\documentclass[]{article} %for 2 columns style without line numbers
\usepackage{graphicx}        % standard LaTeX graphics tool
\newcommand{\be}{\begin{equation}}
\newcommand{\ee}{\end{equation}}

\newcommand{\bea}{\begin{eqnarray}}
\newcommand{\eea}{\end{eqnarray}}

\newcommand{\revision}[1]{#1}

\begin{document}

\begin{titlepage}
    \begin{center}
        \vspace*{1cm}
            
        \LARGE
        \textbf{Power-Law Return-Volatility Cross Correlations of Bitcoin} 
            
        \vspace{0.5cm}
        \LARGE
            
        \vspace{1.5cm}
            
        \textbf{T. Takaishi}
        \vspace{0.5cm}
            
        \large
        Hiroshima University of Economics, Hiroshima 731-0192 JAPAN \\
            
    \end{center}

%\begin{titlepage}
%\title{Power-Law Return-Volatility Cross Correlations of Bitcoin}

%\author{T. Takaishi\thanks{Hiroshima University of Economics, Hiroshima 731-0192 JAPAN 
%PACS {89.65.Gh}{Economics; econophysics, financial markets, business and management} \\
%PACS {05.45.Tp}{Time series analysis} \\
%} }
%

\begin{abstract}
This paper investigates the return-volatility asymmetry of Bitcoin.
We find that the cross correlations between return and volatility (squared return) are mostly insignificant on a daily level.
In the high-frequency region, we find that 
a power-law appears in negative cross correlation between returns and future volatilities,
which suggests that the cross correlation is \revision{long ranged}. 
\revision{We also calculate a cross correlation between 
returns and the power of absolute returns, and} 
we find that the strength of \revision{the cross correlations} depends on the value of the power. 
\end{abstract}

%\maketitle
\end{titlepage}

%\begin{document}

\section{Introduction}
It has long been known that return and volatility are negatively correlated, and 
early studies\cite{Black1976,Christie1982stochastic} attempt to explain the return-volatility asymmetry as a leverage effect:
a drop in the value of \revision{a} stock increases finance leverage or debt-to-equity ratio,
which makes the stock riskier and increases the volatility.
The other promising explanation for the return-volatility asymmetry
is the volatility feedback effect 
discussed in \cite{french1987expected,campbell1992no}: 
if volatility is priced, an anticipated increase in volatility raises the required return, leading to an immediate stock price decline.
Although the two effects suggest the same negative correlations,
the causality is different\cite{bekaert2000asymmetric}.

Comparing the two effects empirically,
Baekaert et al.\cite{bekaert2000asymmetric} and Wu\cite{wu2001determinants} argue that the dominant determinant is the volatility feedback effect.
However, the studies \revision{using} GARCH-type models\cite{Nelson1991Econ,Engle1993JOF,Glosten1993JOF} suggest that
volatility increases more after negative returns than positive \revision{ones},
which favors the leverage effect.

To discuss the full temporal structure of return-volatility asymmetry,
using squared returns as a proxy of volatility,
Bouchaud et al.\cite{bouchaud2001leverage} calculate the return-volatility correlation function and
find that returns and future volatilities are negatively correlated.
On the other hand, reverse correlations, i.e., correlations between future returns and volatilities
are found to be negligible. 
The results are fitted to an exponential function, and 
it is concluded that the correlations are short ranged.
In addition, the decay times\footnote{The correlations are fitted with an exponential function of $\alpha\exp(-t/\tau)$, and the decay time is defined by $\tau$.} 
are estimated to be about \revision{10 (50)} days for stock indices (individual stocks).
 
While for most developed markets, negative correlations between returns and future volatilities are found,
an interesting \revision{phenomenon} is observed in Chinese markets.
Qiu et al.\cite{qiu2006return} 
calculate the return-volatility correlation function for equities in \revision{the} Chinese market and
find that returns and future volatilities are "positively" correlated, which is called the anti-leverage effect.
Further studies\cite{shen2009return,chen2013agent} also support the anti-leverage effect in the Chinese market.

In this study, we focus on the return-volatility asymmetry of the Bitcoin market. 
Since the first proposal of cryptocurrency in 2008\cite{Nakamoto2008},
the Bitcoin system, based on a peer-to-peer network and blockchain technology,
developed quickly, and Bitcoin has become widely recognized as a payment medium.
In recent years, a large body of literature has 
investigated various aspects of Bitcoin, e.g., 
hedging capabilities\cite{dyhrberg2016hedging}, 
inefficiency\cite{urquhart2016inefficiency,bariviera2017some,alvarez2018long,kristoufek2018bitcoin}, multifractality\cite{takaishi2018},
extreme price fluctuations\cite{beguvsic2018scaling}, liquidity and efficiency\cite{wei2018liquidity,takaishi2019market}, 
transaction activity\cite{koutmos2018bitcoin}, complexity synchronization\cite{fang2018multiscale}, long memory\cite{zargar2019long}, and so forth.

Although the return-volatility asymmetry of Bitcoin has been investigated 
using various models, such as asymmetric GARCH-type and stochastic volatility,
it seems that a consistent picture of the return-volatility asymmetry of Bitcoin 
has not yet been obtained.
For instance, while Bouoiyour et al.\cite{bouoiyour2016bitcoin} observe a volatility asymmetry that reacts to negative news rather than positive,
Katsiampa\cite{katsiampa2017volatility} and Baur et al.\cite{baur2018bitcoin} find an inverted volatility asymmetry that reacts to positive news rather than negative.
Moreover, several studies\cite{dyhrberg2016bitcoin,thies2018bayesian,takaishi2018}
find no evidence of a leverage effect in Bitcoin prices. 

Bouri et al.\cite{bouri2016return} investigate return-volatility asymmetry in two periods separated at the price crash of 2013.
They find that, while before the crash Bitcoin shows inverted volatility asymmetry,
after the crash, and for the whole period, no significant volatility asymmetry 
is observed. Using the stochastic volatility model, Philip et al.\cite{phillip2018new} find that one day ahead volatility and returns are
negatively correlated.

Here, we approach the return-volatility asymmetry of Bitcoin
through return-volatility cross correlations. 
We \revision{calculate a cross correlation
between} returns and a power of absolute returns.
This is in part motivated by the existence of the Taylor effect\cite{taylor1986modelling,ding1993long}, 
which suggests that the strength of autocorrelations of a power of absolute returns, $|r|^d$, is
dependent on the value of power $d$,
and, typically, the maximum autocorrelations are obtained 
at $d \approx 1$ for stocks\cite{ding1993long} 
and at $d \approx 0.5$ for exchange rates\cite{genccay2001introduction}.
The Taylor effect is also present for Bitcoin\cite{takaishi2018taylor}.
Thus, we investigate how \revision{the cross correlation} of Bitcoin is dependent on the value of power.

This paper is organized as follows. Section 2 describes the data and methodology.
Section 3 presents the empirical results. Finally, we conclude in Section 4.

\section{Data and Methodology}
We use Bitcoin tick data (in dollars) traded on Bitstamp
from January 10, 2015 to
January 23, 2019 and downloaded from Bitcoincharts\footnote{http://api.bitcoincharts.com/v1/csv/}.
Let $p_{t_i}; t_i= i\Delta t;  i=1,2,...,N$ be the time series of Bitcoin prices
with sampling period $\Delta t$.
We define the return, $R_{i}$, by the logarithmic price difference, namely,
\be
R_{i+1}=\log p_{t_{i+1}} -\log p_{t_{i}}.
\ee
In this study, we consider high-frequency returns with $\Delta t= 2$ min, 
and \revision{we} also consider daily returns.
We further calculate the normalized returns by $r_i=(R_i-\bar{R})/\sigma_R$, where 
$\bar{R}$ and $\sigma_R$ are the average and standard deviation of $R_i$, respectively.
We \revision{calculate the cross correlation}, $CC_d (j)$, between returns 
and \revision{the $d$-th power} of absolute returns at lag $j$ 
as
\be
CC_d(j) = \frac{E[(r_t-\mu_r)(|r_{t+j}|^d -\mu_{|r|^d})]}{\sigma_r \sigma_{|r|^d}},
\ee
where $\mu_r$ and $\mu_{|r|^d}$ are the averages of $r_i$ and $|r_i|^d$,
and $\sigma_r$ and $\sigma_{|r|^d}$ are the standard deviations of $r_i$ and $|r_i|^d$, respectively. 
$E[O_j]$ in Eq.(2) stands for the average over $N-j$ values of $O_j$.
For $d=2$, Eq.(2) reduces to the usual definition of the return-volatility correlation that uses squared returns as a proxy of volatility\cite{shen2009return,chen2013agent}
except the normalization.

We calculate $CC_d (j)$ for $d=0.1$ to $3.0$ every 0.1 step. 
For positive $j$s, $CC_d (j)$ at $d=2$ evaluates the relationships between returns and future volatilities.
The reverse correlations, i.e., relationships between future returns and volatilities, 
are obtained for negative $j$s.

\begin{figure}
%\vspace{5mm}
\centering
\includegraphics[height=5.5cm]{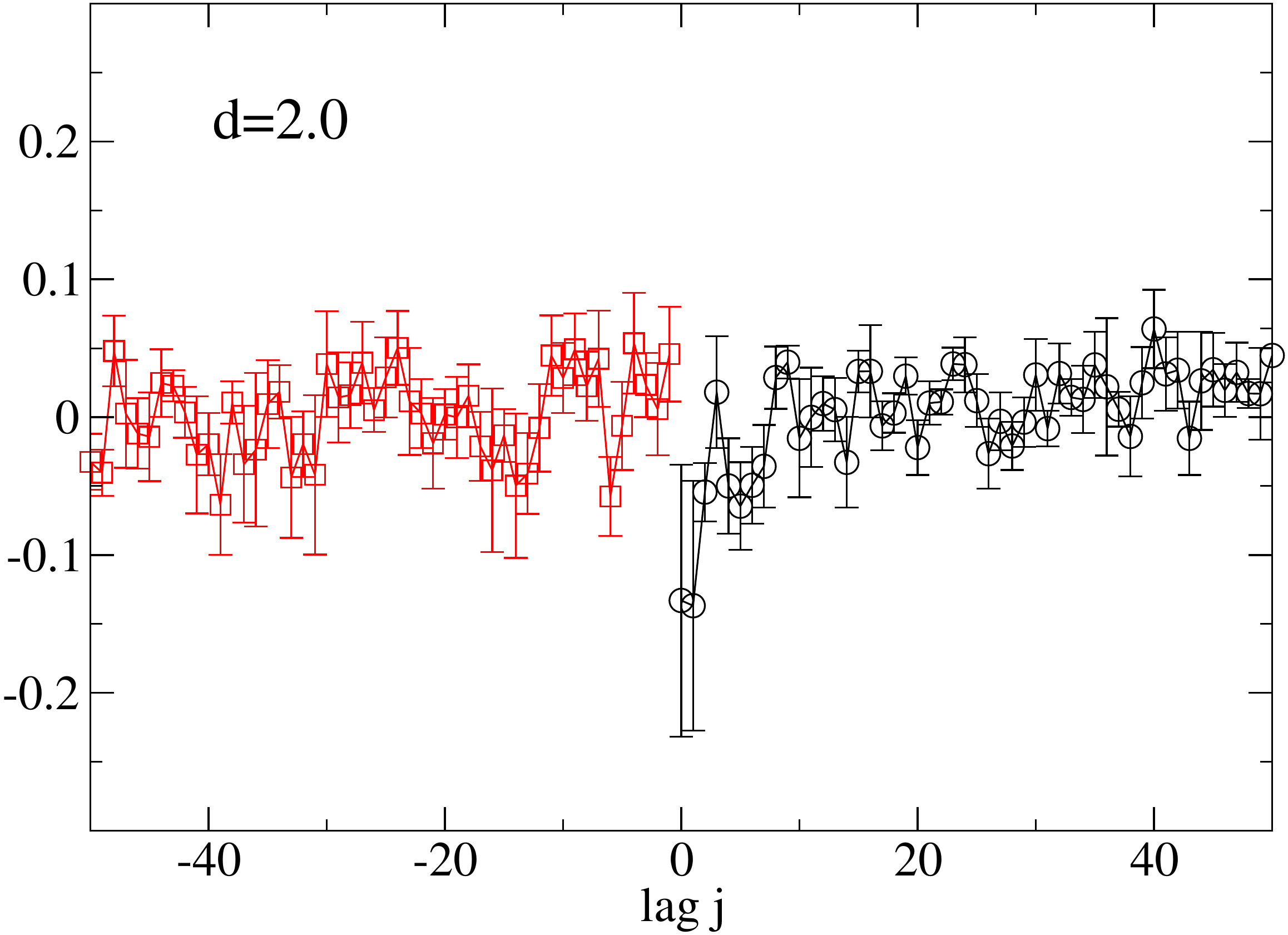}
\caption{
Cross correlation $CC_d(j)$ for daily returns as a function of lag $j$ \revision{at $d=2.0$}.
Error bars of data points represent one sigma errors calculated by the \revision{jackknife} method.
}
%\vspace{-2mm}
\end{figure}

\begin{figure}[ht]
%\vspace{5mm}
\centering
\includegraphics[height=5.5cm]{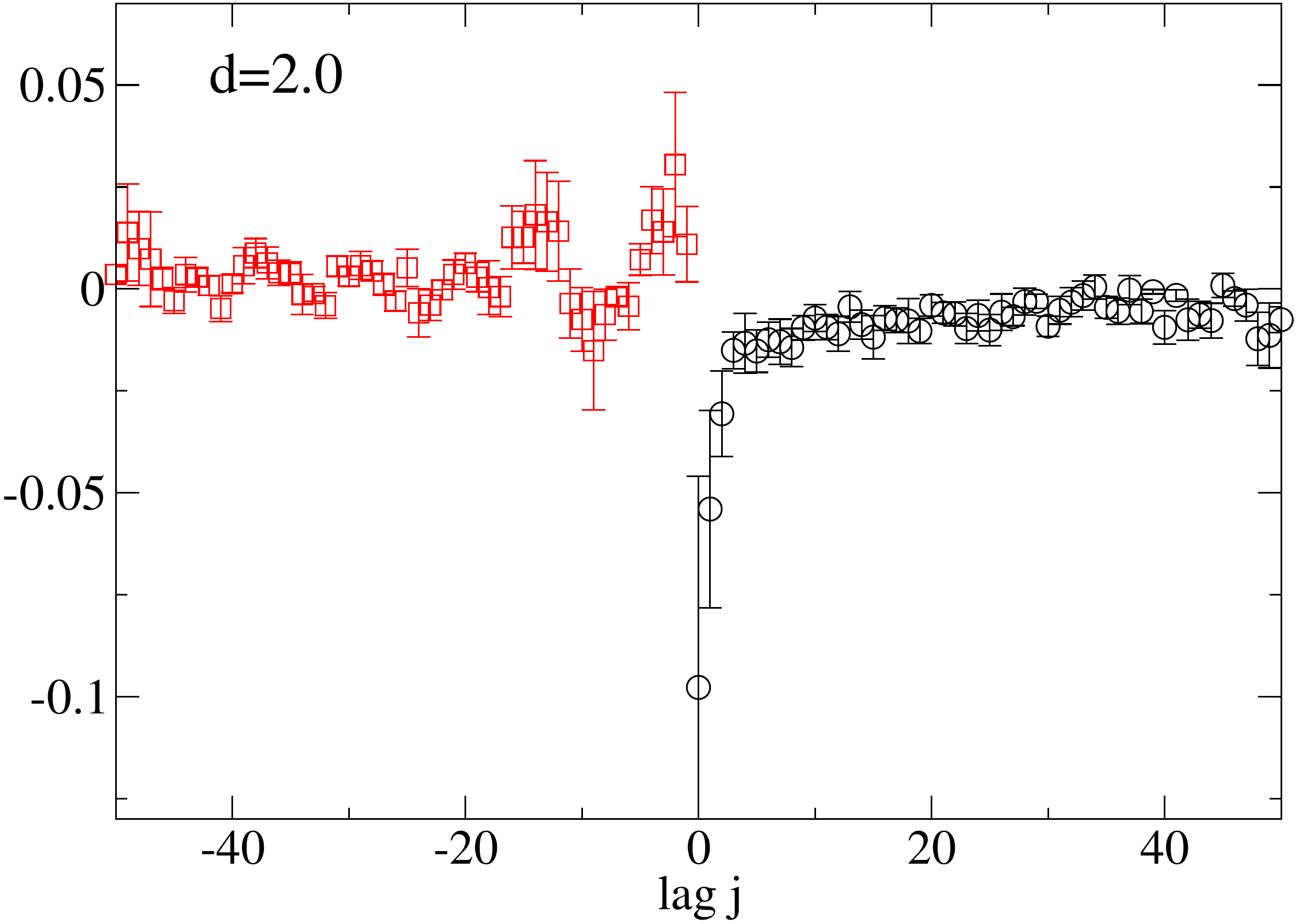}
\caption{
Cross correlation $CC_d(j)$ for high-frequency returns as a function of lag $j$ \revision{at $d=2.0$}.
Error bars of data points represent one sigma errors calculated by the \revision{jackknife} method.
}
%\vspace{-2mm}
\end{figure}

\section{Empirical Results}

First, in Figure 1, we show the cross correlation, $CC_d (j)$, of the daily returns for $d=2.0$.
The cross correlations are mostly consistent with zero for both positive and negative lags, $j$, 
except for $j=0$ and 1, at which negative correlations are observed.
For other $d$s, similar results are obtained.
Thus, at the daily level, the cross correlations are mostly insignificant,
except for contemporaneous and small, positive lags.

\begin{figure}
\includegraphics[height=5.5cm]{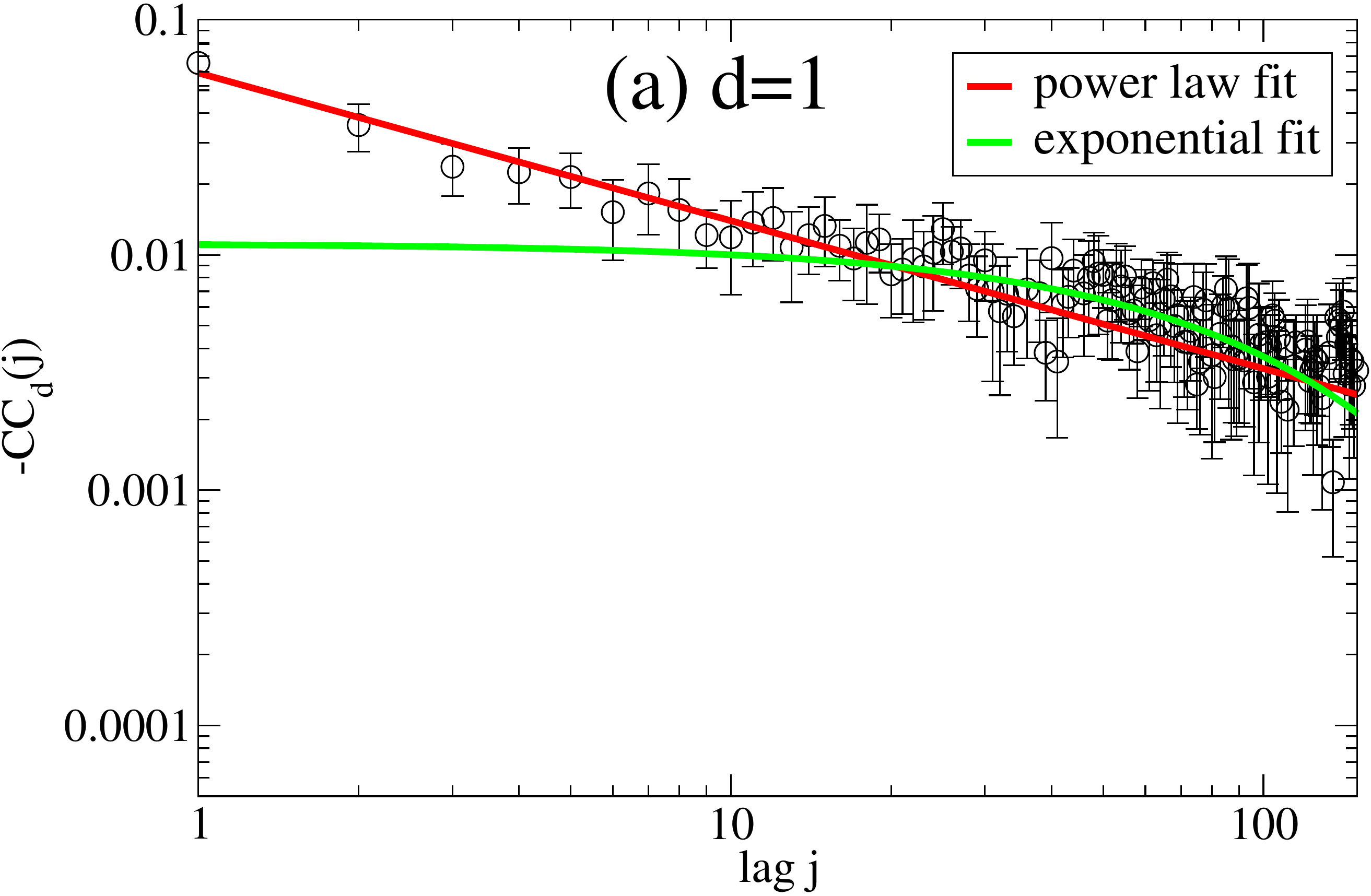}
\includegraphics[height=5.5cm]{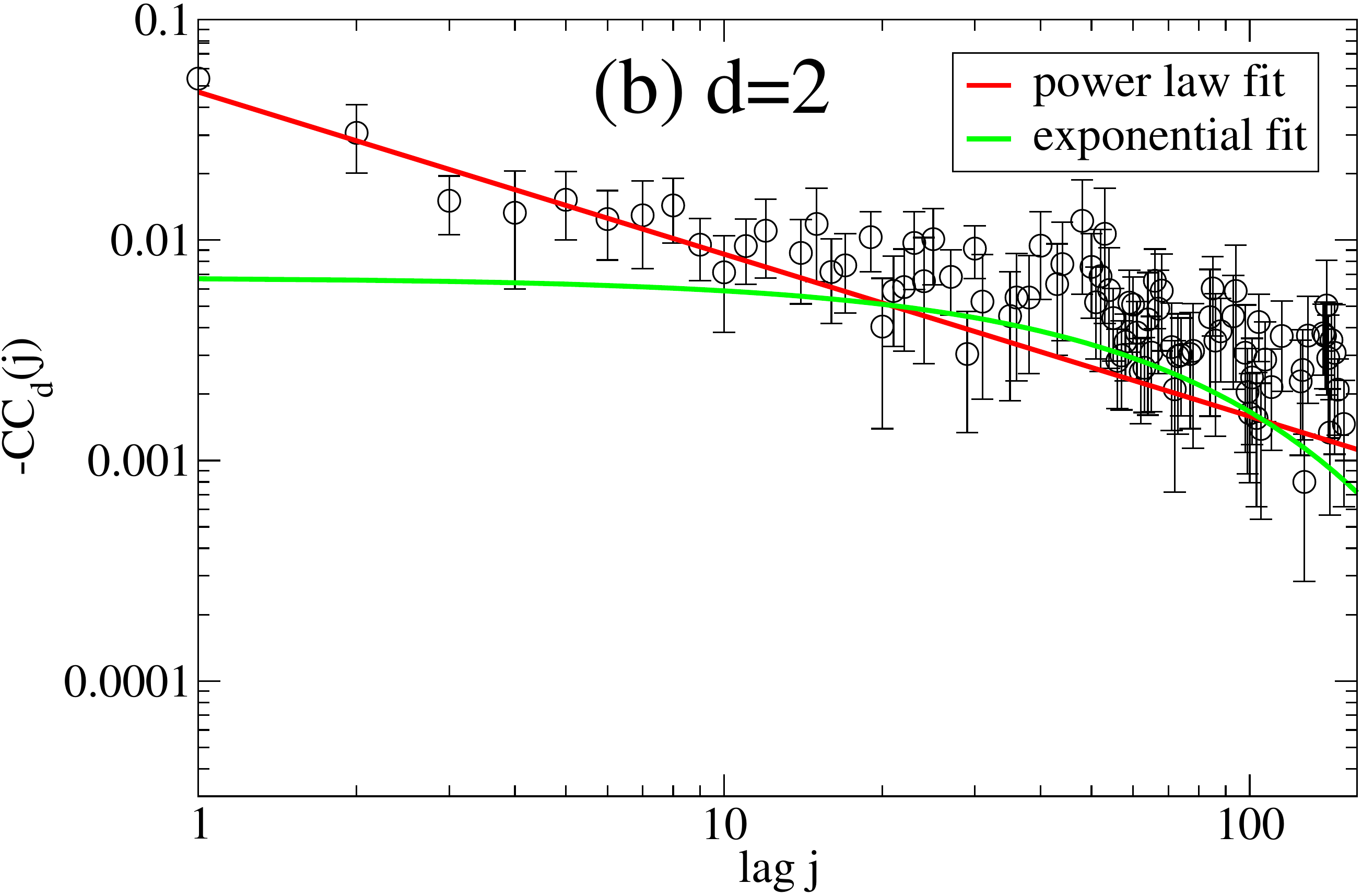}
%%\hspace{-4cm}
\caption{
\revision{Cross correlation} $CC_d(j)$ at $d=1$ and 2 in log-log scale.
(a) d=1 and (b) d=2.
Error bars of data points represent one sigma errors calculated by the \revision{jackknife} method.
\revision{For better visibility, noisy data that are consistent with zero within 1.5 sigma errors are omitted from the pictures.} 
The red (green) solid curve represents the power law (exponential) fitting to the data.
The reduced chi square: (a) 0.796 (power law) and 1.09 (exponential), (b) \revision{1.13 (power law)} and 1.33 (exponential). 
}
%\vspace{-2mm}
\end{figure}

\begin{figure}
%\vspace{-10mm}
\centering
\includegraphics[height=5.5cm]{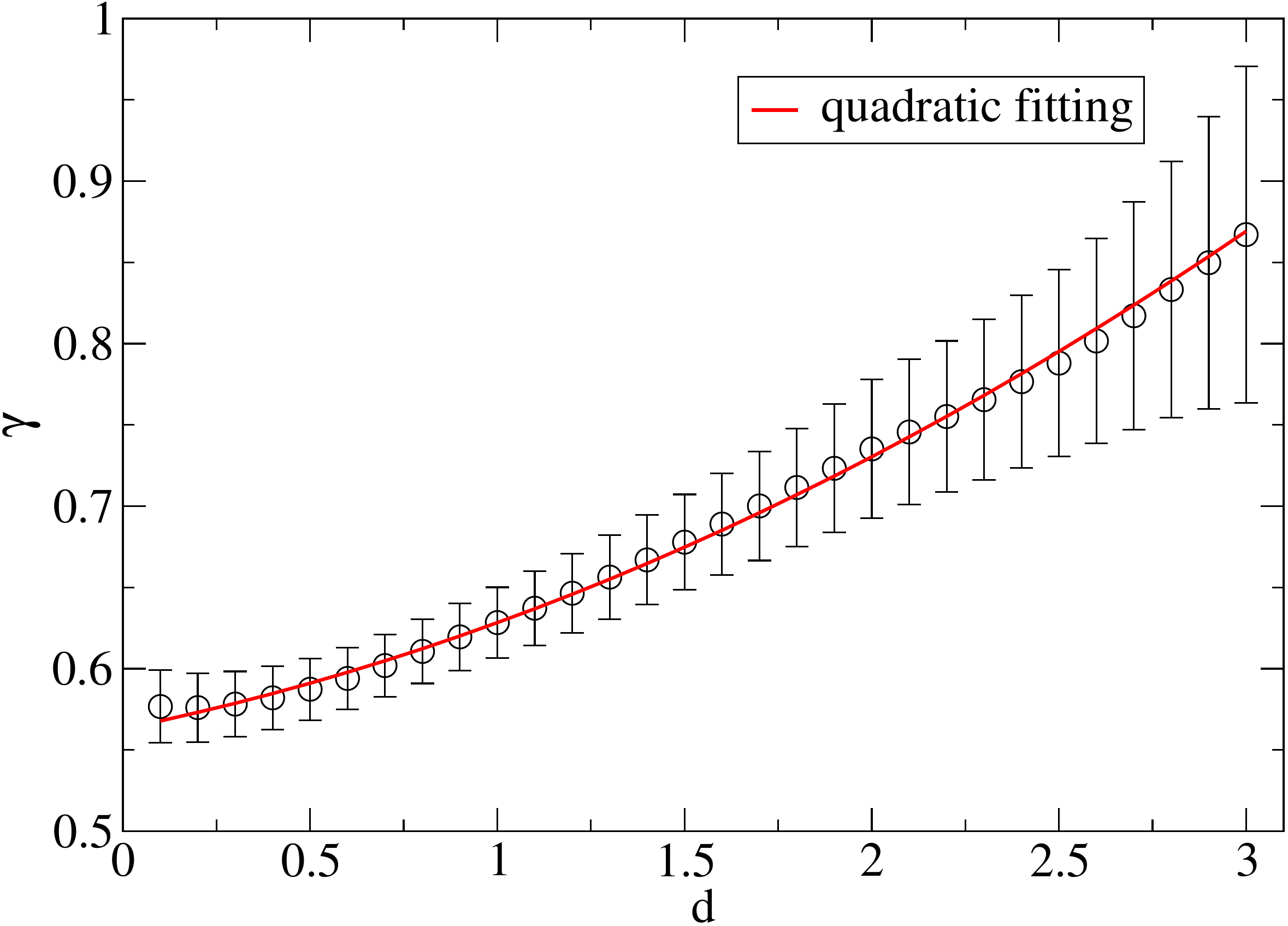}
\caption{
Fitting results of $\gamma$ as a function of $d$, \revision{where $\gamma$ is a parameter of the power law function of $\kappa j^{-\gamma}$}.
Error bars of the results represent the asymptotic standard errors of the parameter.
The solid curve represents the quadratic fitting to $\gamma$.
}
%\vspace{-10mm}
\end{figure}

Next, in Figure 2, we show the cross correlation, $CC_d (j)$, calculated with 2-min, high-frequency returns
for $d=2.0$.
For positive $j$s, we find negative cross correlations lasting from small to large lags,
which is consistent with the results observed for developed markets\cite{bouchaud2001leverage,bollerslev2006leverage}. 
For negative $j$s, we observe positive, but smaller, cross correlations at several small lags. 
For larger (negative) lags, the cross correlations are consistent with zero.
For the contemporaneous correlations, i.e., $j=0$, we observe negative cross correlations.

To examine the scaling properties of the cross correlations 
at positive lags\footnote{Since the cross correlations at negative lags quickly become consistent with zero
at very small lags, 
we only consider those at positive lags.},
we plot negative values of the results, i.e., $-CC_d(j)$ in Figure 3 in log-log scale.

We fit the cross correlations with the power law function of $\kappa  j^{-\gamma}$ and the exponential function of $\alpha \exp(-j/\tau)$
in a range of $j=[1,200]$, where 
$\kappa $,$\gamma$, $\alpha$, and $\tau$ are fitting parameters.
The fitting results of the power law (exponential) function are depicted by the red (green) curve in Figure 3.
We find that the cross correlations are better described by the power law function than by the exponential function. 
\revision{In particular}, we recognize that the exponential function does not adequately describe the data points of cross correlations at small lags.
This finding is different from the results of previous studies that observe exponential behavior in the cross correlation\cite{bouchaud2001leverage,qiu2006return,chen2013agent}.
The exponential behavior in the cross correlation indicates that the cross correlation quickly disappears as the lag increases, i.e.,
the correlation is short ranged. 
On the other hand, the power law behavior\footnote{More precisely, the power law exponent, $\gamma$, should be $\gamma<1$ for a long-range behavior ( e.g., \cite{beran1994}).
In addition, the summation of the correlations diverges for $\gamma<1$.
As seen in Figure 4, the condition of $\gamma<1$ is satisfied, at least for $d<3$.} 
that we observe indicates that the cross correlation decreases slowly with the lag, i.e., 
the correlation is \revision{long ranged}.

\begin{table}
\centering
\caption{
\revision{Results of fitting to a quadratic function, $\gamma(d)= \alpha d^2 + \beta d +\rho $. The values in parentheses represent the asymptotic standard errors of the fitting parameters.}}
%\scriptsize
%\hspace{-10mm}
\begin{tabular}{cccccc}
\hline
                  & $\alpha$   & $\beta$ & $\rho$      \\
\hline
Bitcoin          & 0.0184(13) & 0.0470(35)  & 0.5630(18)  \\
\hline
\end{tabular}
%\vspace{-8mm}
\end{table}

In Figure 4,
we plot the results of $\gamma$ as a function of $d$
and find that $\gamma$ increases with $d$.
We fit the results to a quadratic function, $\gamma(d)= \alpha d^2 + \beta d +\rho $, where $\alpha,\beta$, and $\rho$ are fitting parameters;
the fitting results are listed in Table 1. From the fitting results, we recognize that for $d \rightarrow 0$, the power $\gamma$ 
seems to approach the value around 0.56.
To investigate the strength of the cross correlations, we plot $\kappa$ as a function of $d$ in Figure 5.
\revision{More precisely, $\kappa$ represents the strength of the cross correlations at lag $j=1$.}
We find that $\kappa$ is a convex function and that the maximum strength is obtained around $d\approx 1.4$.
Thus, the correlation $CC_d (1)$ at $d\approx 1.4$ gives a stronger correlation than the traditional cross correlation defined at
$d=2$. 

\begin{figure}[ht]
%\vspace{-5mm}
\centering
\includegraphics[height=5.5cm]{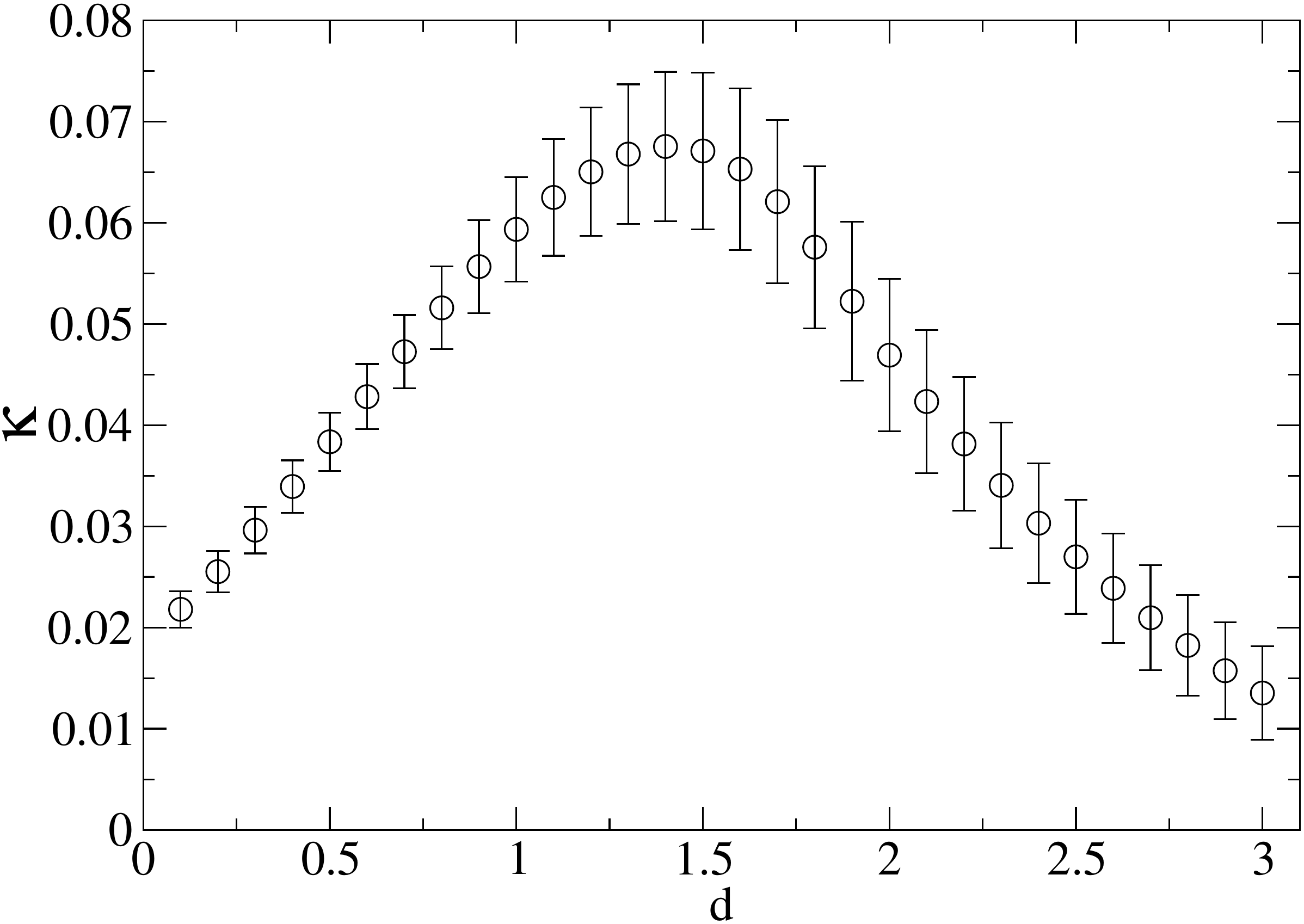}
\caption{
Fitting results of $\kappa$ as a function of $d$, \revision{where $\kappa$ is a parameter of the power law function of $\kappa j^{-\gamma}$.
$\kappa$ corresponds to the strength of a cross correlation at lag $j=1$.}
\revision{The error} bars of the results represent the asymptotic standard errors of the parameter.
}
\vspace{-2mm}
\end{figure}

\section{Conclusion}
At the daily level, cross correlations are mostly insignificant for Bitcoin.
By examining high-frequency Bitcoin returns,
we find that returns and future volatilities are negatively correlated and 
the cross correlations between returns and future volatilities show power law behavior.
We calculate \revision{cross correlations between} returns and 
\revision{the $d$-th power} of absolute returns and 
find that the maximum cross correlation is obtained at $d\approx 1.4$.
Thus, we \revision{were able to} obtain clear evidence on the cross correlation by choosing 
other values of $d$, rather than the traditional value of $d=2$.

Our findings on cross correlations suggest that, 
in modeling asset time series, we should more seriously consider models that produce power law behavior in the cross correlations.

For example, Ref.\cite{roman2008skewness} proposes a fractional random walk model combined with a simple auto-regressive conditional
heteroskedastic model, denoted as FRWARCH, and finds that the FRWARCH model exhibits a power law in the cross correlations.

There exist universal properties, such as volatility clustering and no autocorrelations in returns, that appear across various assets.
These properties are called the stylized facts (e.g., \cite{Cont2001QF}).
The existence of stylized facts suggests that the price formation is governed by \revision{certain} common dynamics.
If Bitcoin \revision{has} a different property in the cross correlation from other assets,
there could exit a different type of dynamics in Bitcoin.
To come to \revision{a definite} conclusion \revision{about} whether the power law behavior only appears in \revision{Bitcoin,} 
it would be desirable to examine other assets \revision{in detail}.

\section*{Acknowledgement}
Numerical calculations for this work were carried out at the
Yukawa Institute Computer Facility and at the facilities of the Institute of Statistical Mathematics.
This work was supported by JSPS KAKENHI \revision{(Grant Number JP18K01556)} and
the ISM Cooperative Research Program (2018-ISMCRP-0006).

\bibliography{mybibfile}

\end{document}